\documentclass[aps,prd,preprint,floats,nofootinbib]{revtex4}

\newcommand{\beq}{\begin{equation}}
\newcommand{\eeq}{\end{equation}}

\begin{document}

\title{The Principle of Mediocrity}

\author{Alexander Vilenkin}

\address{
Institute of Cosmology, Department of Physics and Astronomy,\\ 
Tufts University, Medford, MA 02155, USA}

\begin{abstract}

Recent developments in cosmology suggest that much of the universe is in a state of explosive, accelerated expansion, called inflation.  We live in a "bubble" where inflation has ended, and other bubbles with diverse properties are constantly being formed.   Most, if not all, of these bubbles are beyond our cosmic horizon and cannot be directly observed.  I discuss the origin of this new worldview, its possible observational tests, and its implications for the beginning and the end of the universe.

\end{abstract}

\maketitle

\section*{Cosmic inflation}

The universe as we know it originated in a great explosion that we call the Big Bang.  For nearly a century cosmologists studied the aftermath of this explosion: how the universe expanded and cooled down, how  galaxies were gradually pulled together by gravity, etc.  The nature of the Bang itself came into focus only relatively recently.  It is the subject of the theory of inflation, which was developed in the early 1980's by Alan Guth, Andrei Linde and others and has led to a radically new global picture of the universe.  According to this new picture, remote regions beyond our horizon are strikingly different from what we observe here and may even obey different laws of physics.  Here I will discuss the origin of the new worldview, its possible observational tests, and its implications for the beginning and the end of the universe.

I will start with a brief review of the theory of inflation.  The key role in this theory is played by a peculiar entity called "false vacuum".  Vacuum is just empty space, but according to modern particle physics it is very different from "nothing".  It is a physical object, endowed with energy density and pressure, and can be in a number of different states.  Particle physicists refer to these states as different vacua.  The properties and the types of elementary particles differ from one vacuum to another.  The gravitational force induced by a false vacuum is rather peculiar: it is repulsive.  The higher the energy of the vacuum, the stronger is the repulsion.  The word "false" refers to the fact that this kind of vacuum is unstable.  It decays into a low-energy vacuum like ours, and the excess energy goes to produce a hot fireball of particles and radiation.  I should emphasize that false vacua with these strange properties were not invented for the purposes of inflation; their existence follows from particle physics and General Relativity.

The theory of inflation assumes that at some early time in its history the universe was in the state of a high-energy false vacuum.\footnote{Why was it so?  This is a good question, and I will have something to say about it later in this article.}  The repulsive gravitational force produced by that vacuum would then cause a super-fast, exponential expansion of the universe.  There is a characteristic doubling time in which the size of the universe would double.  This is similar to economic inflation: if the rate of inflation is constant, the prices would double, say, every 10 years.  Cosmic inflation is a lot faster than that: depending on the model, the doubling time can be as short as $10^{-37}$~sec.  In about 330 doubling times the size of the universe will grow by a factor of $10^{100}$.  No matter what its initial size is, the universe will very quickly become huge.
Since the false vacuum is unstable, it eventually decays, producing a hot fireball, and that's the end of inflation.  The fireball continues to expand by inertia and evolves along the lines of standard Big Bang cosmology.  Decay of the false vacuum plays the role of the Big Bang in this scenario.

The theory of inflation explained some otherwise mysterious features of the Big Bang, which simply had to be postulated before.  It explained the expansion of the universe (it is due to the repulsive gravity of the false vacuum), its high temperature (due to the high energy density of the false vacuum), and its observed homogeneity (false vacuum is very homogeneous: apart from quantum fluctuations, it has a constant energy density).  The theory has also made a number of testable predictions.  It predicted that on the largest observable scales the universe should be accurately described by flat, Euclidean geometry.  It also predicted a nearly scale-invariant spectrum of small Gaussian density perturbations. These predictions have been spectacularly confirmed by observations.  By now inflation has become the leading cosmological paradigm.

\section*{Eternal inflation}

Now that the theory of inflation is supported by the data in our observable region, we should have some trust in what it tells us about the big picture --- the structure of the universe beyond our cosmic horizon.

The end of inflation is triggered by quantum, probabilistic processes and does not occur everywhere at once.  Regions where false vacuum decays somewhat later are "rewarded" by a large inflationary expansion, so false vacuum regions tend to multiply faster than they decay.  In our cosmic neighborhood inflation ended 13.7 billion years ago, but it most likely still continues in remote parts of the universe, and other "normal" regions like ours are constantly being formed.  This never-ending process is called {\it eternal inflation}.\footnote{The eternal nature of inflation is not automatic, but it is very generic.  Practically all models of inflation that have been discussed so far predict eternal inflation.}  

The details of false vacuum decay are model-dependent; here, I will focus on models where it occurs through bubble nucleation.  The low-energy regions then appear as tiny microscopic bubbles and immediately start to grow, rapidly approaching the speed of light.  The bubbles keep growing without bound; in the meantime they are driven apart by the inflationary expansion, making room for more bubbles to form.  We live in one of the bubbles and can observe only a small part of it.  No matter how fast we travel, we cannot catch up with the expanding boundaries of our bubble, so for all practical purposes we live in a self-contained bubble universe.  An unlimited number of bubbles will be formed in the course of eternal inflation.  (For a review of inflation, including eternal inflation, see, e.g., Guth \& Kaiser 2005.)

\subsection*{A metaphysical interlude}

At this point I would like to mention a remarkable and, to my mind, somewhat disturbing consequence of this picture of the universe (Garriga \& Vilenkin 2001).  Since the number of bubble universes is unlimited, and each of them expands without bound, they will contain an unbounded number of regions of the size of our horizon. In each of these regions, the initial conditions at the big bang are set by random quantum processes during inflation, so all possible initial conditions will be realized with some probability.

Now, the key point is that the number of distinct states in which any such region can be is finite.  How is this possible? I can, for example, move my chair by one centimeter, by half-centimeter, by a quarter centimeter, and so on, and apparently I have an infinite number of possible states right there --- because I can move it by an infinite number of possible displacements which get smaller and smaller. However, states that are too close to one another cannot be distinguished, even in principle, due to the quantum uncertainty. So quantum mechanics tells us that the number of distinct states (in a finite volume) is finite. The number of quantum states in our observable region has been estimated as ${\cal N}\sim \exp(10^{122})$.  This is an unimaginably large number. But the important point is that it is not infinite.

Thus, we have a finite number of states occurring in an infinite number of regions. The inevitable conclusion is that each state having a non-zero probability occurs an infinite number of times. In particular, there is an infinite number of earths identical to ours. This means that scores of your exact duplicates are now reading this sentence. There should also be regions with all possible variations. For example, there are some regions where the name of your dog is different and others where dinosaurs still roam the earth.\footnote{You may be wondering whether all this is happening at the same time.  This question does not have a definite answer, since time and simultaneity are not uniquely defined in General Relativity.  If for example we use matter density as the time variable in a bubble universe, then at each moment of time the bubble interior is an infinite hyperbolic space, and each of us has an infinite number of duplicates presently living in our bubble.}  

Note that infinity of space (or time) is not by itself sufficient to warrant these conclusions.  We could, for example, have the same galaxy endlessly repeated in an infinite space.  So we need some "randomizer", a stochastic mechanism that would pick initial states for different regions from the set of all possible states.  Even then, the entire set may not be exhausted if the total number of states is infinite.  So the finiteness of ${\cal N}$ is important for the argument.  In the case of eternal inflation, the finiteness of ${\cal N}$ and the randomness of initial conditions are both guaranteed by quantum mechanics.

\section*{The multiverse}

So far I assumed that other bubble universes are similar to ours in terms of their physical properties.  But this does not have to be so.  String theory, which is at present our best candidate for the fundamental theory of nature, admits an immense number of solutions describing vacua with diverse physical properties.  These solutions are characterized by different compactiÞcations of 
extra dimensions, by branes wrapped around extra dimensions in different ways, etc. The number of possibilities is combinatorial and can be as high as $10^{500}$ (Lerche et al. 1987, Bousso \& Polchinski 2000).  Each solution corresponds to a vacuum with its own types of elementary particles and its own values for the constants of Nature.

Now combine this with the theory of inflation.  Wherever they occur in the universe, high-energy vacua will drive exponential inflationary expansion.  Transitions between different vacua will occur through bubble nucleation, so there will be bubbles within bubbles within bubbles.  Each bubble type has a certain probability to form in the inflating space.  So inevitably, an unlimited number of bubbles of all possible types will be formed in the course of eternal inflation.

This picture of the universe, or multiverse, as it is called, explains the long-standing mystery of why the constants of nature appear to be fine-tuned for the emergence of life (see, e.g., Linde 1990).  I will give you just one example: the neutron mass.  In our universe neutrons are slightly heavier than protons.  An isolated neutron decays into a proton, an electron, and an antineutrino, but neutrons bound in atomic nuclei are stabilized by nuclear forces.  Suppose now that we decrease the neutron mass by 1\%. Then neutrons would become lighter than protons, and this would allow protons to decay into neutrons and lighter particles. As a result the atomic nuclei will lose their electric charge. So there will be nothing to keep electrons in atoms, and they will fly away. Thus, if the neutron mass is decreased by 1\%, we end up in a universe without atoms, and it is hard to imagine how life, which is anything like ours, could exist in such a universe.

Now, if we increase the mass of the neutron by 1\%, it becomes so massive that it decays even inside a nucleus, turning into a proton. The electric repulsion between protons will then tear the nucleus apart, and the only possible atom will be a single proton combined with an electron, which is hydrogen. Once again, it is hard to see how life can be possible in a universe with no chemical elements other than hydrogen.

The situation is similar with other constants. If you change them by relatively small amounts, you end up with a universe that is not fit for life. This seems to suggest that the constants were fine-tuned by the Creator, in order to make a bio-friendly universe for us to live in. This is exactly what the advocates of intelligent design have been telling us all along!

The multiverse picture offers a different explanation.  The constants of nature take a wide range of values, varying from one bubble to another.  Intelligent observers exist only in those rare bubbles in which, by pure chance, the constants happen to be just right for life to evolve. The rest of the multiverse remains barren, but no one is there to complain about that. (For a non-technical review of multiverse ideas, see Vilenkin 2006, Susskind 2006, Greene 2011.)

Some of my colleagues find the multiverse theory alarming.  Any theory in physics stands or falls depending on whether its predictions agree with the data.  But how can we verify the existence of other bubble universes?  Some distinguished cosmologists, like Paul Steinhardt and George Ellis, have even argued that the multiverse theory is unscientific, because it cannot be tested, even in principle. 

Surprisingly, observational tests of the multiverse picture may in fact be possible.  One possibility is to look for observational signatures of bubble collisions.  As it expands, our bubble will occasionally collide with other bubbles.  In fact, it will experience an infinite number of collisions in the course of its history.  Each such collision will produce an imprint in the cosmic microwave background (CMB) --- a round spot of higher or lower radiation intensity (Aguirre \& Johnson 2011, Kleban 2011).   The CMB polarization within the spots is also predicted to have a characteristic pattern (Czech et al. 2011).   A detection of a spot which has the predicted characteristics and stands significantly above the background would provide direct evidence for the existence of other bubble universes. The search is now on (Feeney et al. 2011), but unfortunately there is no guarantee that a bubble collision has occurred within our cosmic horizon.  Hence, failure to find signatures of bubble collisions on the microwave sky cannot be regarded as evidence against eternal inflation.

Another interesting possibility is that our bubble universe could have tunneled out of an inflating vacuum where some of our three spatial dimensions were compactified.  One can then hope to detect some residual asymmetry in the expansion rate or in the spectrum of CMB temperature fluctuations.  In a simple model with one initially compact dimension this effect appears too small to be detected (Blanco-Pillado \& Salem 2010, Graham et al. 2010), but in other models the situation may be more favorable.  

\section*{The Principle of Mediocrity}

As in a criminal trial, in the absence of direct evidence for the multiverse, one can look for indirect, or circumstantial evidence.  The idea is to use our theoretical model of the multiverse to predict the constants of nature that we can expect to measure in our local region.  One selection criterion is the so-called Anthropic Principle, first introduced by Brandon Carter (1974).  There are many different formulations of this principle in the literature, but most people understand it as stating the obvious fact that we can expect to measure only such values of the constants that are consistent with the existence of life.   This "principle", however, is guaranteed to be true, so it is not very useful for testing the theory.  

In order to make testable predictions, we have to take a somewhat different approach (Vilenkin 1995). We can use the theory to derive the probability distribution for the constants measured by a randomly picked observer in the multiverse.  Assuming that we are typical observers --- the assumption that I called the Principle of Mediocrity --- we can then predict the expected range of values for the constants in our bubble.  The width of this range will depend on the confidence level  at which we want to make the prediction.  For example, if the desired confidence level is 95\%, we should discard 2.5\% at both tails of the distribution.\footnote{Similar ideas have been suggested by Gott (1993), Leslie (1989), Page (1996) and Bostrom (2002).  I should add that Carter's own interpretation of the Anthropic Principle was close to the Principle of Mediocrity.}

This strategy has been applied to the energy density of our vacuum $\rho_v$, also known as Òdark energyÓ. Steven Weinberg (1987) (see also Linde 1987) has noted that in regions where $\rho_v$ is large, it causes the universe to expand very fast, preventing mater from clumping into galaxies and stars. Observers are not likely to evolve in such regions.  Values of $\rho_v$ much smaller than needed for galaxy formation require unnecessary fine-tuning and are also rather unlikely.  Calculations showed that most galaxies (and therefore most observers) are in regions where the dark energy density is about the same as the density of matter at the epoch of galaxy formation.  The prediction is therefore that a similar value should be observed in our part of the universe (for a review and references see Vilenkin 2007).

For the most part, physicists did not take these ideas seriously, but much to their surprise, dark energy of roughly the expected magnitude was detected in astronomical observations in the late 1990Õs.  As of now, there are no alternative explanations for the observed value of $\rho_v$. This may be our first evidence that there is indeed a huge multiverse out there. It has changed many minds.

\subsection*{Are we really typical?}

The Principle of Mediocrity has been a subject of much controversy.  It asserts that we are typical observers, but there will always be some unfortunate creatures in the multiverse who will measure atypical values of the constants.  How can we be sure that we are not them?  Hartle \& Srednicki (2007) have argued, for example, that we should never assume ourselves to be typical in some class of observers, unless we have evidence to back up that assumption.  The Principle of Mediocrity makes an opposite claim --- that we should assume ourselves to be typical in any class that we belong to, unless there is some evidence to the contrary (Garriga \& Vilenkin 2008).

Actually, I am surprised that this issue is so controversial, since one can easily convince oneself that the Principle of Mediocrity provides a winning betting strategy.  I will illustrate this with a simple example.  Imagine that as you arrive to a meeting of the Royal Society, the organizers put a white or black hat on you.  They have removed all mirrors, so you don't know the color of your hat.  You notice though that 80\% of people around you wear white hats and 20\% wear black hats.  There may or may not be some system as to how the hats are distributed.  For example, the color could be correlated with your sex, age, hight, etc. --- but you don't know.

Now, in order to register for the meeting, you have to bet 100 pounds on the color of your hat.  How are you going to bet?  One strategy is to assume that you are typical among the participants and bet that your hat is white.  Another approach is to say that you don't really know whether you are typical or not.  Then you throw a coin and bet at random.  With the first choice, 80\% of people will win, while with the second choice only 50\% win.  Clearly, the Principle of Mediocrity provides a better betting strategy.  With more information, you can improve your odds by narrowing your reference class accordingly.  For instance, if you are a woman and you notice that most women wear black hats, you should bet that your hat is black.

\subsection*{The measure problem}

A more serious challenge for the theory of the multiverse is the so-called measure problem.
As we discussed earlier, any event having a nonzero probability will happen in the course of eternal inßation, and it will happen an inÞnite number of times. Statistical predictions are based on relative frequencies of events in the limit of $t \to\infty$. One finds however that the outcome sensitively depends on the limiting procedure.  More precisely, it depends on what variable we use as time $t$.  One possible choice is the "proper time" measured by the clocks of comoving observers.  Another natural choice is the expansion factor (or scale factor) of the universe.  The crux of the problem is that the volume of an inflating universe grows exponentially with time, and the numbers of all kinds of events grow accordingly.  As a result, most of the events will always be near the cutoff time, so it is not surprising that the resulting probability measure depends on exactly how the cutoff is introduced.   

On the positive side, predictions for the CMB multipoles and for the dark energy are not very sensitive to the choice of measure.  But as a matter of principle, the theory will remain incomplete until the measure is fully specified.  

The measure problem has been around for nearly two decades.  During this time, a number of different measures have been proposed and their properties have been investigated (for a recent review see Freivogel 2011).  This work has shown that some of the proposals lead to paradoxes or to a conflict with the data and should therefore be discarded.  For instance, the proper time measure performed rather poorly, while the scale factor measure is still in the running.  It is unlikely, however, that this kind of phenomenological analysis will yield a unique prescription for the measure.  This suggests that some important element may be missing in our understanding of cosmic inflation.  

Some people feel the problem is so grave that it puts the validity of the theory of inflation seriously in doubt (e.g., Steinhardt 2011).  But this is the view of only a small minority of cosmologists.  Personally, I think the situation with the theory of inflation is similar to that with Darwin's theory of evolution some 100 years ago.  Both theories greatly expanded the range of scientific inquiry, proposing an explanation for something that was previously believed impossible to explain.  In both cases, the explanation was very compelling, and no viable alternatives have been suggested.  Darwin's theory was widely accepted, even though some important aspects remained unclear before the discovery of the genetic code.  The theory of inflation may be similarly incomplete and may require additional new ideas.  But it also has a similar air of inevitability.

\section*{Beginning and end of the universe}

If inßation has no end, could it also have no beginning? This would allow us to avoid many 
perplexing questions associated with the beginning of the universe.  Once you have a universe, its evolution is described by the laws of physics, but how do you describe the beginning? What caused the universe to appear? And who sets the initial conditions for the universe? It would be an attractive solution if we could say that the universe has always been in the state of eternal inflation, without a beginning and without end.

This idea, however, runs into an unexpected obstacle. Arvind Borde, Alan Guth and I (2003) have proved a theorem, which says that even though inflation is eternal to the future, it cannot be eternal to the past\footnote{More precisely, the theorem says that all geodesics in an inflationary spacetime, except a set of measure zero, are incomplete to the past.} --- which means that inflation must have had some sort of a beginning. We are then faced with the question of what happened before inflation. And whatever the answer is, we can keep asking: ÒAnd what happened before that?Ó So it seems that one of the most basic questions of cosmology --- What was the beginning of the universe? --- does not have a satisfactory answer.
 
The only way around this problem of infinite regress that has been suggested so far is the idea that the universe could be spontaneously created out of nothing. We often hear that nothing can come out of nothing. Indeed, matter has positive energy, and energy conservation demands that any initial state should have the same energy. However, it is a mathematical fact that the energy of a closed universe is equal to zero.  In such a universe, the positive energy of matter is exactly compensated by the negative energy of the gravitational field, so the total energy is zero. Another conserved quantity is the electric charge, but once again it turns out that the total charge must vanish in a closed universe.  This is not difficult to understand.  Suppose the universe has the form of a 3-dimensional sphere, and imagine placing a positive charge at the "South pole" of that sphere.  The lines of force emanating from the charge will then converge at the North pole, indicating that there must be an equal negative charge there.  Thus, you cannot add an electric charge to a closed universe without adding an opposite charge someplace else.  

If all the conserved numbers of a closed universe are equal to zero, then there is nothing to prevent such a universe from being spontaneously created out of nothing. In quantum mechanics, any process which is not strictly forbidden by the conservation laws will happen with some probability. 
The newly born universes can have a variety of sizes and can be filled with different types of vacua. Analysis shows that the most probable universes are the ones having the smallest initial size and the highest vacuum energy (for more discussion see Vilenkin 2006, Ch. 17). Once the universe is formed, it starts rapidly expanding, due to the high energy of the vacuum. This provides the beginning for the scenario of eternal inflation.
 
You can ask: What caused the universe to pop out of nothing? Surprisingly, no cause is needed. If you have a radioactive atom, it will decay, and quantum mechanics gives the decay probability in a given interval of time. But if you ask why the atom decayed at this particular moment and not the other, the answer is that there is no cause: the process is completely random. Similarly, no cause is needed for quantum creation of the universe.

I would like to close this section with some important news about the end of the world. It is 
often said that if the dark energy is a cosmological constant, then the universe will continue expanding forever. This is true for our bubble universe as a whole, but not for our local region. In 
the multiverse picture, there must be a large number of negative-energy vacua, and bubbles 
of such vacua will inevitably form within our (nearly zero-energy) vacuum. At some point, 
probably in a very distant future, our neighborhood will be engulfed by a negative-energy 
bubble. The expansion will then locally turn into contraction, and our region will collapse 
to a big crunch. The bubble will arrive without warning, since it expands at nearly the speed of light. In fact, it may be rushing towards us this very moment.

\section*{Outlook}

In summary, I have described the new worldview that has emerged from inflationary cosmology.  
According to this view, inflation is a never ending process, constantly producing new "bubble universes" with diverse properties.  This multiverse picture can be tested both by direct observation of bubble collisions and indirectly, using the Principle of Mediocrity.  The prediction for the dark energy based on this principle has already been confirmed. Here I will mention some other observational tests that have been suggested in the literature. 

A potentially testable feature of bubble universes is their negative spatial curvature.  The curvature parameter $\Omega_k$ is different for different bubbles, depending on the amount of inflation in the bubble interior. The probability distribution for $\Omega_k$ has been studied by Freivogel et al. and by De Simone \& Salem (2010).  They found that the detectable range of values for the curvature $(|\Omega_k| \gtrsim 10^{-4})$  has a non-negligible probability, but at the same time the broad tail of the distribution extends to values that are too small to be detected.  Apart from the curvature,
quantum fluctuations in the parent vacuum may also produce a characteristic feature in the spectrum of gravitational waves inside the bubble.  Detection of either of these effects would provide additional evidence for eternal inflation.  

The Principle of Mediocrity has also been applied to explain the amount of dark matter in the universe.  The composition of dark matter is unknown, and one of the best motivated hypotheses is that it is made up of very light particles called axions.  The density of axionic dark matter is set by quantum fluctuations during inflation and varies from one place in the universe to another.
Its value affects the formation of galaxies; hence, there is an anthropic selection effect.  Tegmark et al. (2006) (see Linde 1988 for earlier work) have calculated the resulting probability distribution and found that the observed value of the dark matter density is close to the peak of the distribution.  If indeed the dark matter turns out to be axionic, this can be counted as a success of the theory.

Multiverse predictions for the neutrino masses have been worked out by Tegmark et al. (2005) and Pogosian et al. (2004), with the conclusion that the sum of neutrino masses should be $\sim 1 ~eV$.  It is intriguing to note that recent neutrino oscillation experiments, as well as cosmological data, point to the existence of sterile neutrinos with $m\sim 1 ~eV$ (e.g., Hamann et al. 2010).

A major unresolved problem of the inflationary cosmology is the measure problem.  Its resolution may require radically new ideas.  One possibility that has been recently suggested (Garriga \& Vilenkin 2009) is that the dynamics of the inflationary multiverse has a dual, "holographic" description, in the form of a quantum field theory defined at the future boundary of spacetime.  The measure of the multiverse can then be related to the short-distance cutoff in that theory.  
This and other possibilities are now being explored.

\section*{Acknowledgement}

This work was supported in part by grant PHY-0855447 from the National Science Foundation.

\end{document}